\documentclass{article}

\usepackage{arxiv}

\usepackage[utf8]{inputenc} % allow utf-8 input
\usepackage[T1]{fontenc}    % use 8-bit T1 fonts
\usepackage{hyperref}       % hyperlinks
\usepackage{url}            % simple URL typesetting
\usepackage{booktabs}       % professional-quality tables
\usepackage{amsfonts}       % blackboard math symbols
\usepackage{nicefrac}       % compact symbols for 1/2, etc.
\usepackage{microtype}      % microtypography
\usepackage{lipsum}		% Can be removed after putting your text content
\usepackage{graphicx}
\usepackage{natbib}
\usepackage{doi}
\usepackage{amsmath}
\usepackage{algorithm}
\usepackage{algpseudocode}
\usepackage{float}
\usepackage{todonotes}
\usepackage{caption}
\usepackage{subcaption}

\title{Optimizing Transition Strategies for Small to Medium Sized Portfolios}

\author{Nakul Upadhya  \\
	Department of Mechanical and Industrial Engineering\\
	University of Toronto\\
	Toronto, ON, M5S 2E4\\
	\texttt{nakul.upadhya@mail.utoronto.ca} \\
	\And
	Alexandre Granzer-Guay \\
	Department of Mechanical and Industrial Engineering\\
	University of Toronto\\
	Toronto, ON, M5S 2E4\\
	\texttt{alexandre.granzerguay@mail.utoronto.ca} \\
}

\renewcommand{\shorttitle}{Optimizing Small Portfolio Transitions}

\hypersetup{
pdftitle={Optimizing Transition Strategies for Small to Medium Sized Portfolios},
pdfsubject={q-bio.NC, q-bio.QM},
pdfauthor={Nakul Upadhya, Alexander Granzer-Guay},
pdfkeywords={Portfolio Optimization, Integer Programming, Machine Learning},
}

\begin{document}
\maketitle

\begin{abstract}
This work discusses the benefits of constrained portfolio turnover strategies for small to medium-sized portfolios. We propose a dynamic multi-period model that aims to minimize transaction costs and maximize terminal wealth levels whilst adhering to strict portfolio turnover constraints. Our results demonstrate that using our framework in combination with a reasonable forecast, can lead to higher portfolio values and lower transaction costs on average when compared to a naive, single-period model. Such results were maintained given different problem cases, such as, trading horizon, assets under management, wealth levels, etc. In addition, the proposed model lends itself to a reformulation that makes use of the column generation algorithm which can be strategically leveraged to reduce complexity and solving times. All of the discussed experiments and presented results can be reproduced using our code at \href{https://github.com/upadhyan/Portfolio-Changeover-Optimization}{github.com/upadhyan/Portfolio-Changeover-Optimization}.    
\end{abstract}

% keywords can be removed

\section{Introduction}

Many retail investor strategies found in the literature, make certain assumptions for convenience. In fact, it is most common to ignore a non-partial constraint and fixed transaction costs in lieu of convexity and near optimality. Though reasonable when the portfolios in question boast wealth levels well over \$1 Million, they start breaking down when wealth levels start decreasing to ``young'' or ``novice'' investor levels (below \$500 000). In this paper, we propose to explore this fact through a common and simple strategy. Suppose, an investor currently holds a diversified portfolio. The market experiences a regime switch and the investor responds by converting her portfolio either to a risk-averse strategy or an aggressive strategy to capitalize on upward trends. Such a portfolio conversion may incur significant trading fees and value if the conversion is not executed correctly. We propose a multi-period dynamic model that finds the optimal trades to execute for a prescribed portfolio conversion over a given time horizon. In doing so, we consider small-budget investors and the constraints that come with lower wealth levels, namely, transaction fees, timing, and simplicity.

The presented strategy is a multi-period mixed-integer model that creates a trading plan for a prescribed horizon time frame. At each period, within the horizon, the strategy is recalculated and reassessed for changes in market behavior. Given that the strategic plan relies on a valuable forecast, it's important to adapt to any unforeseen behavior. We demonstrate the effectiveness of our proposed method through a set of numerical experiments using real-world financial data and scenarios. Our results show that our strategy outperforms classical approaches by successfully minimizing trading costs and maximizing portfolio value.

This paper is separated into 6 key sections. We first introduce the literary background that motivates our thesis. Then, we present a succinct problem statement before introducing the reader to the primary component of our model - the multi-period transition framework. Finally, we present our experimental methodology in detail before presenting compelling numerical results which support our proposal. We conclude this paper with thoughts and comments on the observed performance and, we close the discussion with possible avenues for future work.

\section{Related Works}

Most classical work in portfolio allocation theory revolves around algorithms that have certain pragmatic benefits while also possessing some numerical advantages [\citeyear{markowitz1952, sharpe1964, rockafellar2000}]. When such advantages are not present - non-convex models, cardinality models, mixed-integer models, to name a few - the approach will either involve a problem relaxation of sorts or a reasonable assumption. In the case of models that consider transaction costs, until more recently, the empirical method was to ignore or ``convexify'' transaction costs. Though some will ignore transaction costs altogether, the status quo seems to be the use of a convex formulation of transaction costs [\citeyear{mansini2015, lobo2007, boyd2017}]. Such methods consider a linear function for the transaction costs, usually based on trading volume, and embed it into the formulation as a constraint or a Lagrangian relaxation, like in \citet{boyd2017}. 
Alternatively, fixed or fixed and linear transaction costs are considered. In such cases, the authors, \citet*{lobo2007}, argue that this leads to a difficult combinatorial formulation, albeit, with a desirable sparse trading behavior. The nature of the trading constraint promotes the use of fewer trades but involves complicated permutations, in particular, in the case of large portfolios. This is discussed in greater detail in the cited manuscript. Later, we demonstrate how we can circumvent usual relaxations of the combinatorial problem by employing a column generation algorithm fit to our particular problem.

In addition to transaction cost constraints, we choose to constrain our framework to trading non-partial shares. Portfolio weights are traditionally computed in reference to the total portfolio value. In practice this means there would be some deviation between the prescribed weights and the equivalent shares to be bought. In large scale portfolio, a difference of $\pm1$ share in a stock won't affect the relative holding weight of that stock in the portfolio. On the other hand, for small to medium sized portfolio, adding or removing a share may drastically alter the relative holding weight of this same position. As a result, we cannot assume the equivalency between a portfolio with fractional shares and one without and, instead, must enforce a whole share policy. To the best of our knowledge this is an unusual assumption that has not led to any major publication \citep{wickremasinghe2022}.

The main thesis of this paper revolves around a case where an investor must convert a portfolio A to a portfolio B under some prescribed constraint, namely, budget constraints and time constraints. The common approach is to use a continuously moving target and moving current portfolio, as is done in \citet{smith1967}. The future portfolio is determined using a portfolio allocation optimization model and the current portfolio is optimized to aim for the future portfolio while reducing transaction costs. This same approach is also explored in a probabilistic manner, where \citet{collin-dufresne2020}, makes use of a stochastic model to find the optimal trading plan (future portfolio) to adjust between different market regimes. There, the authors, are considering models with wealth levels above 2 million and assessing the benefits and costs of continuously adapting a portfolio to the underlying regime. \citet*{garleanu2013} demonstrate that, in dynamic environments, there is a benefit in employing a rolling horizon method when attempting to aim for a target portfolio. Though there approach focuses on a continuously varying target portfolio, we will demonstrate that this claim also applies to our use case. The referenced literature makes use of larger sized portfolios and methods that consider, at most, one period in the future. Though this is perfectly reasonable, we choose to focus on smaller portfolios in addressing the transition problem, as opposed to the optimization and transition problem. This distinction is important as we believe that most retail investors are rarely rebalancing their portfolios. In fact, we maintain that such investors do not have the time or resources to continuously assess the optimality of their portfolio and therefore attribute a low level of importance to portfolio optimization. Instead, they tend to prefer minimal friction (losses) when they do elect to rebalance or adjust their portfolios. We present this thesis clearly in the next section.

\section{Formal Problem Definition}

The principle idea behind the portfolio turnover strategy is simple when we consider the turnover problem over a single period. Given a set of stocks $A$ and a starting portfolio $P_0$, we aim to find a set of stock trades $Z$ (which can be split into purchases $z^+$ and sells $z^-$) to transition our holdings to one that satisfies a target set of assets $\mathcal{T}$.  In such a scenario the goal is to minimize transaction costs and maximize portfolio value while converting the portfolio to its target. This portfolio value is equivalent to the combined value of the stocks in the portfolio plus any leftover cash at the final time step $C$. 

As we are aiming to replicate the environment faced by retail traders, there are a few problem constraints imposed. For one, each time we buy or sell any asset, we must pay a trading fee $F$. Additionally, the transition must be made without borrowing cash (all $C$ values must be positive) or short-selling (portfolios must be non-negative at each time step). 

The portfolio turnover problem increases in difficulty when we extend the trading window from a single period to instead consider $\mathbf{T} >> 1$ periods. The trader now can examine the varying asset prices across time $Y_t$ and identify opportune times to execute trades to ensure minimal portfolio value is lost while still attaining the target state. However, the true prices are unknown for any future period and instead, the trader must instead rely on forecasted prices $\hat{Y_t}$, which are often very error-prone. This means that the increased opportunities for portfolio growth in a multi-period transition come with equivalent risk. As such, it is useful to have a framework that allows a trader to specify the level of risk they are comfortable with and provides a trading strategy that accommodates their preference. 

\section{Transition Framework}
Throughout this paper, we will make reference to a multi-period model called a ``Receding Horizon'' (RH) model - popularized in finance by \citet*{boyd2017}. The model is explained in detail in the cited work and thus, will only be briefly covered in this section. The premise of the model is to leverage a rolling horizon approach over a finite horizon. In other words, the framework finds a desired plan of actions for a given number of future periods - Lookahead - and continuously recalculates this plan based on new observed data. The key insight is that the plan is calculated using data-driven forecasting methods (Section 5.1). The algorithm is presented below (Algorithm \ref{alg:MPTF}). 
\begin{algorithm} 
\caption{Multi-Period Transition Framework}\label{alg:MPTF}
\begin{algorithmic}[H]
\Require $P_0, \mathcal{T}, Y_{-L:0}, \mathbf{X}_{-L:0}, V_0, \mathbf{T}, C_0$, OPT, Forecaster, MarketObserver
\Ensure $Y_0^T\mathcal{T} +  C_0 \leq V_0$
\State $t \gets 0$
\State $P_t \gets P_0$
\State $C_t \gets C_0$
\State $Y \gets Y_{-L:t}$
\While{$t \leq \mathbf{T}$}
    \State $V_t \gets Y_t^TP_t + C_t$ \Comment{Calculate portfolio value at current timestep}
    \State $\hat{Y}_{t+1:\mathbf{T}}\gets \text{Forecaster}(Y_{-L:t}, \mathbf{X}_{-L:t})$ \Comment{Predict unknown time steps with known information}
    \State $Z_{t:\mathbf{T}}, W_{t:\mathbf{T}} \gets \text{OPT}([Y_t\ \hat{Y}_{t+1:\mathbf{T}}], P_t, C_t, \mathcal{T})$ \Comment{Run Optimization Policy}
    \State $P_{t+1} \gets P_t + Z_t$ \Comment{Execute trades at current timestep}
    \State $C_{t+1} \gets C_t - Y_t\cdot Z_t - F\Vec{1} \cdot W_t$    \Comment{Calculate cash}
    \State $Y_{t+1} \gets \text{MarketObserver}()$ \Comment{Observe the market for new price information}
    \State $t \gets t + 1$
\EndWhile
\Ensure $P_\mathbf{T} \geq \mathcal{T}$
\end{algorithmic}
\end{algorithm}

At each timestep, we utilize known price information and external covariates ($\mathbf{X}$) to predict future price information. This information is passed into an integer programming based optimization policy which calculates a potential set of trades for the remaining periods. From the calculated buys and sells, we execute the trades prescribed for the current period ($Z_t$) and pay the required fees for these transactions ($W_t$). We then observe the changes that occur in the market, updating the value of our portfolio ($V_t$) based on new price information and repeating this process till the end of our trading horizon. 

Rolling trades out iteratively can help an investor anticipate price movements that may impact one's ability to liquidate or acquire certain position \citep{boyd2017}. Additionally, including the new price information can help correct errors in our forecast that result from incorrectly predicting the trend or unexpected market events. These factors combined help minimizes losses when performing a portfolio conversion.

\subsection{Multi-Period Trading Policies}

While multiple policies can be plugged into our portfolio transition framework, we propose various formulations that work well in this environment, all of which are variations of the following formulation: 

\begin{align}
\text{\text{OPT}}(Y, P_t, C_t, \mathcal{T}) = \underset{z,w}{\text{maximize}}&\ C_\mathbf{T} +  Y_{\mathbf{T}}^TP_\mathbf{T} - \sum_{\tau = t}^\mathbf{T}\sum_{a \in A} F(w^+_{\tau, a} + w^-_{\tau,a})&\ \label{eqn:gen_obj}\\
\text{subject to}&\ P_{\tau} = P_{\tau-1} + z^+_\tau - z^-_\tau &\ \forall \tau = t+1, \ldots, \mathbf{T} \label{eqn:price_update}\\ 
&\ C_\tau = C_{\tau-1} - Y_\tau^T(z_\tau^+ - z_\tau^-) - \sum_{a\in A}F(w^+_{\tau,a} + w^-_{\tau,a}) &\ \forall \tau =  t+1, \ldots, \mathbf{T} \label{eqn:cash_constraint}\\ 
&\ z^+_\tau \leq M_\tau w^+_\tau &\ \forall \tau = t, t+1, \ldots, \mathbf{T} \label{logic1}\\
&\ z^-_\tau \leq M_\tau w^-_\tau &\ \forall \tau = t, t+1, \ldots, \mathbf{T} \label{logic2} \\
&\ P_\tau \geq \Vec{0} &\ \forall \tau = t, t+1, \ldots, \mathbf{T} \label{short}\\ 
&\ C_\tau \geq 0 &\ \forall \tau = t, t+1, \ldots, \mathbf{T} \label{leverage}\\ 
&\ P_\mathbf{T} \geq \mathcal{T} \label{target}\\
&\ w^+_\tau + w^-_\tau \leq 1&\ \forall \tau = t, t+1, \ldots, \mathbf{T} \label{ve} \\
&\ w_\tau^+, w_\tau^- \in \{0,1\}^{|A|} &\ \forall \tau = t, t+1, \ldots, \mathbf{T} \\ 
&\ z_\tau^+, z_\tau^- \in \mathbb{Z}_+^{|A|} &\ \forall \tau = t, t+1, \ldots, \mathbf{T}
\end{align}
In this formulation, $z^+_\tau$ and $z^-_\tau$ are vectors that represent the magnitude of buys and sells suggested at time step $\tau$ and $Z_\tau = z^+_\tau - z^-_\tau$. To model trading fees, we include two binary logical vectors $w^+_\tau$ and $w^-_\tau$ that represent whether a buy or sell is executed and $W_\tau = w^+_\tau + w^-_\tau$. The magnitude and logical vectors are linked to each other via constraints \ref{logic1} and \ref{logic2}. For both of these constraints, $M_\tau$ is an upper bound on the number of shares that can be bought or sold during each time step $\tau$. More formally, 
\begin{align}
    U_t &= V_t& \\
    U_\tau &= V_t\prod_{i = t+1}^\mathbf{\tau}\max\{1 + \frac{Y_{i,a} - Y_{i-1,a}}{Y_{i-1,a}}\ \forall a \in A\}\ &\forall \tau = t+1, \ldots, \mathbf{T} \\ 
    M_\tau &= \lceil \frac{U_{\tau}}{\min{Y_{\tau}}} \rceil& \forall \tau = t, t+1,\ldots, \mathbf{T}
\end{align}
Where \(U_t\) represents the upper bound of the portfolio value at time \(t\) and \(V_t\) the portfolio value at time \(t\). The intuition behind this bound is based on the perfect trade assumption. In order to maximize portfolio value at each time step, a trader would move the entire value of their portfolio into the asset with the largest positive price gain between the current period and the next. The upper bound of the accessible value at each time step is then equivalent to the initial portfolio value times the product of the maximum possible gains at each time step. Using this value, we can then divide this buy the smallest priced asset at each $\tau$ to represent the maximum number of assets that could be transacted in that period, which is equivalent to $M_\tau$.

Constraints \ref{eqn:price_update} and \ref{eqn:cash_constraint} are linear expressions that model the held assets and cash at each time step in our horizon. As previously mentioned, the goal of our algorithm is to transition to a portfolio that satisfies a target portfolio $\mathcal{T}$. This target portfolio represents the minimum number of each stock we must have in our final portfolio, and this requirement is captured in constraint \ref{target}. Additionally, we represent the requirements prohibiting short selling and leverage in constraints \ref{short} and \ref{leverage} respectively. 

Constraint \ref{ve} is a constraint introduced to reduce the number of effectively symmetric points and in turn reduce the solver runtime. Without this constraint, solutions, where we buy and sell the same asset at the same timestep, are permitted, resulting in a large feasible space with multiple solutions that result in the same net trade. However, this action will never be optimal due to the presence of trading costs, and constraint \ref{ve} trims these possibilities by constraining the model to take a single action with a given asset at each timestep, either buy or sell but not both. 

For our objective, we aim to maximize our end portfolio value while also minimizing the amount wasted on trading fees (\ref{eqn:gen_obj}). One may notice that the impact of trading fees is included both explicitly in the objective function (\ref{eqn:gen_obj}) and through the cash relations between times (\ref{eqn:cash_constraint}). This double inclusion is intentional as we want our optimization to emphasize reducing wasted money. 

This base formulation can also be thought of as a "day trading" formulation where the recommendations will be extremely aggressive. 

\subsubsection{Directional Trading Formulation}
In our first formulation (also known as the directional formulation), only trades that move the portfolio towards the target are permitted. This means that any trades that attempt to generate profit are prohibited and the main purpose of this policy is to simply minimize the cost of the transition across our time period.  

To implement this, we can split $A$ into 2 sets at each time step the optimization is run on based on whether we are allowed to buy that stock ($A^+_t := \{a \in A; P_{t,a} < \mathcal{T}_a \}$) or sell that stock ($A^-_t := \{a \in A/A^+_t\}$). We can then add the following constraints on $w^+$ and $w^-$ to our base formulation:

\begin{align}
    w^+_{\tau, a} \leq 0 &\ \forall \tau = t, t+1, \ldots, \mathbf{T}\ \forall a \in A^-_t \label{buy_restrict}\\ 
    w^-_{\tau, a} \leq 0 &\ \forall \tau = t, t+1, \ldots, \mathbf{T}\ \forall a \in A^+_t \label{sell_restrict}
\end{align}

One note is that $A$ must be re-partitioned in each iteration of the multi-horizon transition framework. As all trades are executed with imperfect future information, the optimal set of trades may include purchasing more stock than is necessary to meet the target as the forecast for that asset may indicate a strong future performance. However, if the forecast was incorrect and the value of the stock instead falls drastically over multiple periods, a locked $A^+$ and $A^-$ set will force the trader to accept these losses, reducing the benefits of rolling trades out gradually. By redistributing the assets between the buy and sell sets as new information arrives, we allow for course corrections and loss prevention, making our approach more robust to market shocks.

The addition of constraints \ref{buy_restrict} and \ref{sell_restrict} also remove the need for constraint \ref{ve} as now each asset can only be bought or sold in all time steps.

\subsubsection{Penalized Directional Trading Formulation}
While the directional trading policy allows for a safe transition to the target, the policy does not subscribe trades that do not directly move towards the target portfolio. This can result in lost profit opportunities as some of these superfluous trades could take advantage of price fluctuations to generate extra cash which may be used to move our portfolio closer to satisfying the target while reducing the number of assets that need to be sold. 

However, we also want to be retain some of the benefits of allowing trades in only a single direction, namely the reduction in volatility that comes from limiting the number of trades. To balance these issues, we introduce our Directional Penalty formulation. Instead of hard constraining trades that move us away from our portfolio, we instead implement a soft constraint that instead penalize our objective by the value of the trades that move in the wrong direction by adding the following term to the objective function of our base formulation:
\begin{align}
    -\lambda \sum_{\tau = t}^{\mathbf{T}}[\sum_{a \in A^+_t}Y_{\tau,a}z^-_{\tau,a} + \sum_{a \in A^-_t}Y_{\tau,a}z^+_{\tau,a}]    
\end{align}

Instead of prohibiting trades that move us away from the target, we instead penalize our objective by a portion of the value of those trades, encouraging the algorithm to only recommend trades if the profit potential is sufficiently high. As we show in our experimental results, this formulation has the benefit of allowing users to specify the risk they are comfortable with by increasing $\lambda$ for less risk and decreasing it for more risk. When $\lambda = 0$, this formulation is identical to the base formulation and is an extremely aggressive day trading formulation.

\subsection{Naive Trading}
As a comparison baseline of the multi-period trading policies, we introduce a naive policy where trading is only permitted on the first day, therefore the target portfolio must be satisfied on this day. The formulation for this policy is as follows:

\begin{align}
    \underset{z,w}{\text{maximize}}&\ Y_0^TP + C - \sum_{a \in A} F(w^+_{a} + w^-_{a})\\
    \text{subject to}&\ P = P_0 + z^+ - z^- \\
    &\ C = C_0 + Y_0^T(z^--z^+) - \sum_{a \in A} F(w^+_{a} + w^-_{a}) \\ 
    &\ z^+ \leq Mw^+ \\
    &\ z^- \leq Mw^- \\ 
    &\ P \geq \mathcal{T} \\
    &\ P \geq 0 \\ 
    &\ C \geq 0 \\ 
    &\ z^+,z^- \in \mathbb{Z}^{|A|} \\
    &\ w^+, w^- \in \{0,1\}^{|A|}
\end{align}

For this formulation, $M = \lceil (Y_0^TP_0 + C_0) / \min{Y_0}\rceil$ or the value of the portfolio divided by the price of the least valuable asset. The purpose of this formulation is to solely to transition to a satisfactory portfolio in the smallest number of trades possible and by solving this formulation, we are able to receive an estimate of the number of trades required to transition from the initial portfolio to one that satisfies the target for any given portfolio. 

Additionally, we can track the resultant portfolio generated by this optimization and evaluate how it performs during and after our trading window to evaluate the utility of the multi-period policies.  

\section{Methodology}
To evaluate our policies, we generated 254 different scenarios using using historical stock price data from the Toronto Stock Exchange (TSX) from 1999 to 2022. In each scenario, we chose 20 to 50 different stocks and randomly created initial and target portfolios by first deciding a budget between \$15,000 and \$350,000 and then randomly picking stocks from our chosen set until the budget is exhausted. Each scenario was also assigned a fixed trading fee between \$2 and \$9. 

We then ran our transition framework on each instance to simulate how our approach would perform given real market conditions. While the trading fees and portfolio values were varied, the length of the trading horizon ($\mathbf{T}$) was kept constant for every scenario and set to 30 business days. At the end of each simulation, we measured the runtime required to solve our optimization formulations, the percent change in the portfolio value between the first and last periods of the simulation, the number of trades executed, and the total value lost to trading fees. 

\subsection{Forecasting}
While any forecasting method can be used to predict future stock prices, we chose to use N-HiTS \citep{N-HiTS}. N-HiTS is a deep-learning architecture that leverages residual connections, multi-rate signal sampling, and interpolation mechanisms to analyze patterns at different time scales and iteratively construct the end forecast \citep{N-HiTS}. The architecture of N-HiTS is organized into blocks and stacks, with stacks containing multiple blocks and each block consisting of a few fully connected layers that predict weights on a basis vector. For our experiments, we utilized the implementation provided by Darts \citep{Darts}, a user-friendly Python library for time-series forecasting. 

Each scenario has its own instance of the model which was trained for 125 epochs using 5 years of historical data from the start of the training period with a Mean Absolute Percent Error loss function. The model took in 48 days of historical price information and interest rate history to predict price of the stocks over the next 30 days. Each instance had 3 total stacks, 2 blocks per stacks, and 4 fully connected linear layers of 128 width per block. The rest of the hyperparameters such as the learning rate, activation function, and dropout were kept to the defaults provided by Darts. Since each scenario had a different training set, the performance of the forecasts also varied from scenario to scenario. To mitigate this, the average MAPE of the generated forecasts were recorded and scenarios with a MAPE higher than 10\% were excluded from experimentation as often these corresponded to unusual events in the market such as earnings releases, governmental policies, and worldwide pandemics. While this exclusion is not valid in practice, the focus of this work is on the integer programming framework rather than developing a strong forecasting algorithm since there is a large amount of work in this field already \citep{GandhamaSurvey, Survey1}. On average the mean average percent error of our forecasts is approximately 8\%. 
\subsection{Environment}
All experiments were run on a device with a 12th Gen Intel(R) Core i7-12700 @ 2.10 GHz and 32 GB of memory. All the forecasting models were trained using a NVIDIA GeForce RTX 3080. Our optimization models were implemented using Gurobi's Python API \citep{gurobi}.   

\section{Experimental Results}

\subsection{Percent Change Analysis}
Looking at the statistics on the the percent change in portfolio value (Figure \ref{fig:GainStats}), we can see a clear difference in the performance of our various optimization policies. 

\begin{figure}[H]
    \centering
\begin{tabular}{lrrrrr}
\toprule
     policy &  Mean &  Std. Dev. &  Median &   Max &    Min \\
\midrule
      Naive &  1.14 &       3.44 &    0.88 & 11.27 & -10.35 \\
Directional &  1.25 &       3.22 &    1.04 & 11.84 &  -8.29 \\
DirP\_500 &  1.31 &       3.43 &    0.95 & 14.51 & -10.83 \\
  DirP\_75.0 &  1.39 &       3.91 &    0.86 & 30.08 & -11.15 \\
  DirP\_50.0 &  1.47 &       4.26 &    0.97 & 23.62 & -12.07 \\
  DirP\_25.0 &  2.01 &       7.55 &    1.46 & 58.36 & -34.55 \\
     DirP\_0 &  3.18 &      12.59 &    1.62 & 58.59 & -36.50 \\
\bottomrule
\end{tabular}

    \caption{Summary Statistics on the Percent Change (\%) in Portfolio Value}
    \label{fig:GainStats}
\end{figure}

On average, every one of our policies had higher portfolio gains than the naive baseline, clearly showing the value of trading over multiple periods. 
Additionally, we can see that our non-restricted base formulation (DirP\_0) on average almost tripled the expected returns of the Naive formulation. 
However, this increase in average return can be explained by the extremely high standard deviation in the observed changes. 
We can observe from looking at the other Directional Penalty Formulations (DirP\_500, DirP\_75, DirP\_50, and DirP\_25) that our penalty factor $\lambda$ (equal to 500\%, 75\%, 50\%, and 25\% respectively) is inversely correlated to the deviation in returns with higher penalties resulting in a lower return but also lower deviation in returns. In order words, our formulation allows investors to represent their acceptable risk in the problem via the penalty parameter $\lambda$ allowing them to balance the trade off between profit-generation and safe transitions. Additionally, if an investor is extremely risk averse, they can instead opt to use the alternative, pure directional formulation. This formulation not only has a higher average return than the naive formulation, but also a lower deviation therefore lower risk. 

These insights are made more clear when performing a win-loss-tie analysis of the different policies against each other. 

\begin{figure}[H]
    \centering
\begin{tabular}{llllllll}
\toprule
{} & Directional &     DirP\_0 &  DirP\_25.0 &  DirP\_50.0 &  DirP\_75.0 &   DirP\_500 &      Naive \\
\midrule
Directional &         --- &  117/137/0 &  104/150/0 &  114/136/4 &  113/137/4 &  116/135/3 &  144/110/0 \\
DirP\_0      &   137/117/0 &        --- &  129/122/3 &  139/114/1 &  138/116/0 &  138/116/0 &  137/117/0 \\
DirP\_25.0   &   150/104/0 &  122/129/3 &        --- &   146/99/9 &  140/107/7 &  144/101/9 &  151/102/1 \\
DirP\_50.0   &   136/114/4 &  114/139/1 &   99/146/9 &        --- &   81/81/92 &   95/87/72 &  142/112/0 \\
DirP\_75.0   &   137/113/4 &  116/138/0 &  107/140/7 &   81/81/92 &        --- &   92/89/73 &  144/109/1 \\
DirP\_500    &   135/116/3 &  116/138/0 &  101/144/9 &   87/95/72 &   89/92/73 &        --- &  144/110/0 \\
Naive       &   110/144/0 &  117/137/0 &  102/151/1 &  112/142/0 &  109/144/1 &  110/144/0 &        --- \\
\bottomrule
\end{tabular}

    \caption{Win Loss Tie Comparison of Different policies (Row Win/Row Loss/Tie)}
    \label{fig:wlt}
\end{figure}

This analysis shows us that all of our policies have higher returns than the naive formulation most of the time, but the ratio of wins to losses between remains relatively similar across all the formulations, showing that the main difference between the various formulations is not whether they win over the naive formulation, but rather the magnitude of their win. 

Another trend highlighted by this analysis as well as the summary statistics in figure \ref{fig:GainStats} is the diminishing returns from increasing $\lambda$. The number of ties between $\lambda = 25\%$ and $\lambda = 50\%$ is 9, but the number of ties between $\lambda = 75\%$ and $\lambda = 500\%$ is 73, despite the later having over a 400\% difference in value. This is supported by the difference in average gain where the increasing lambda by $25\%$ with low values results in an average decrease of .54\% but increasing it by 25\% from 50\% or by 425\% from 75\% both only result in an .08\% decrease in value. Additionally, as $\lambda$ increases, the behavior converges closer to the directional formulation and in practice, if users specify their safety factor to be incredibly high, any tool leveraging our method should convert that to instead utilize the directional formulation. 

One large factor impacting the observed returns of our policies is the market conditions at the time of the simulation. As mentioned before, the performance of the naive baseline provides information on the how are target portfolio performs at the end of the trading horizon, helping us understand whether or not the optimization was done in a bear or bull market with respect to $A$. 

\begin{figure}[H]
     \centering
     \begin{subfigure}[b]{0.7\textwidth}
         \centering
         \includegraphics[width=\textwidth]{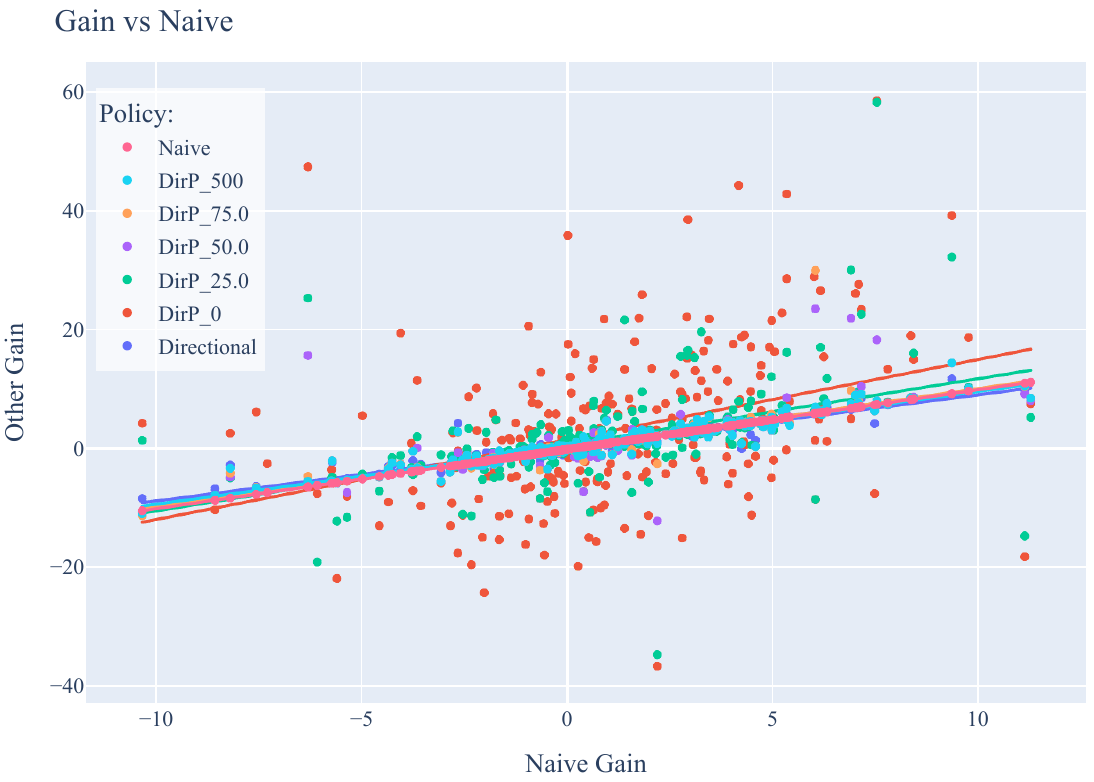}
         \caption{All Policies}
         \label{fig:y equals x}
     \end{subfigure}
     \hfill
     \begin{subfigure}[b]{0.48\textwidth}
         \centering
        \includegraphics[width=\textwidth]{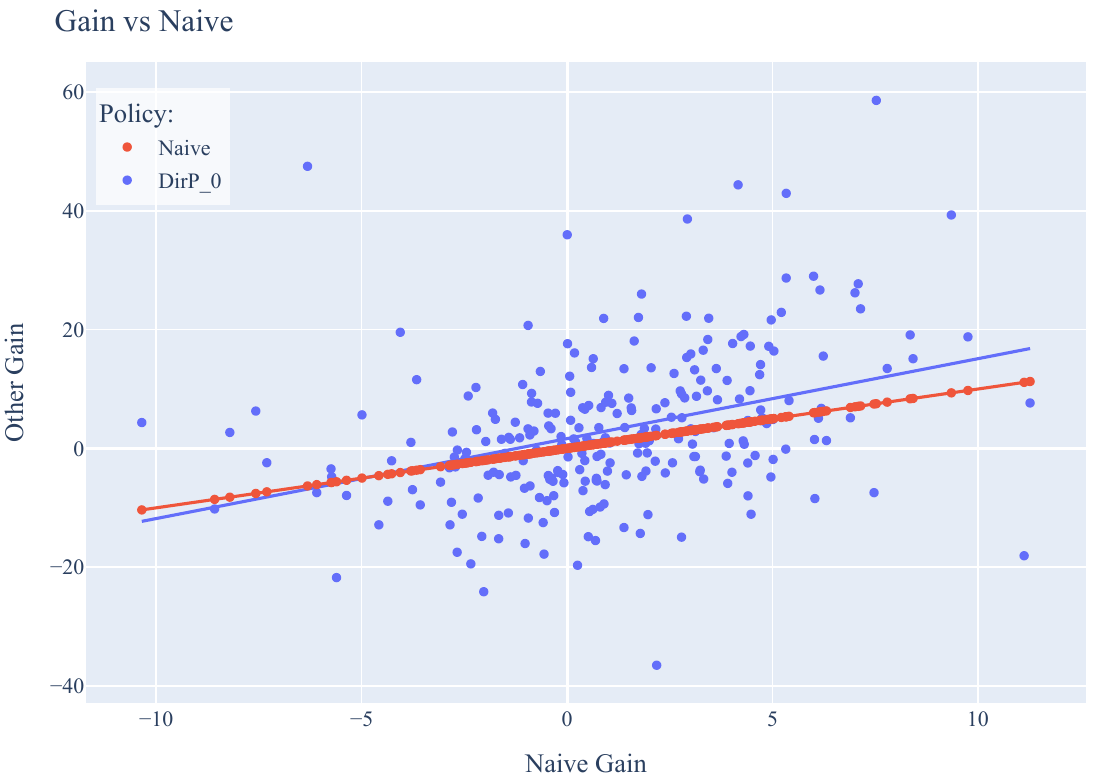}
         \caption{Day Trading / Unpenalized}
         \label{fig:DayVNaive}
     \end{subfigure}
     \hfill
     \begin{subfigure}[b]{0.48\textwidth}
         \centering
        \includegraphics[width=\textwidth]{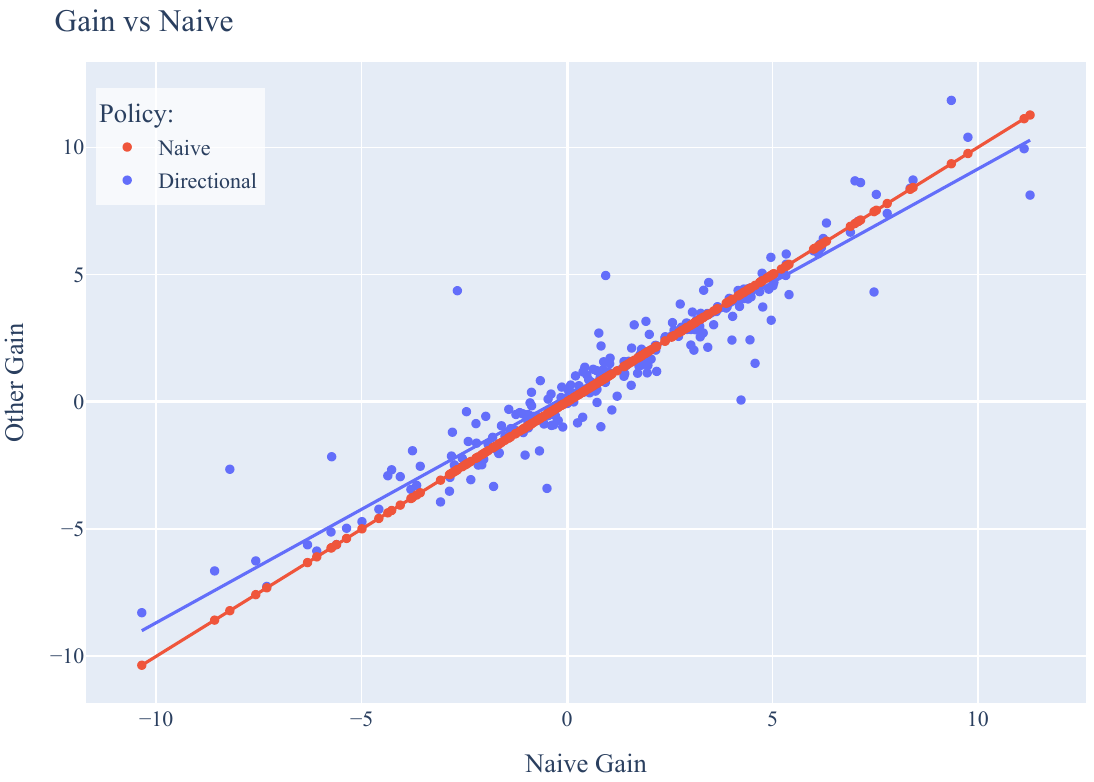}
         \caption{Direction Restricted}
         \label{fig:DirVNaive}
     \end{subfigure}
        \caption{Comparison of Returns to the Naive Baseline}
        \label{fig:NBBaselineComp}
\end{figure}

Looking at the returns of our formulations with respect to market conditions (Figure \ref{fig:NBBaselineComp}), we can see that our different formulations have clear advantages depending on whether or not we are in a bear market. In this figure, the pink line represents the naive performance, or the baseline market condition. 

When examining the performance our unpenalized formulation (Figure \ref{fig:DayVNaive}), we can see that that the trendline fit to the data is steeper than that of the expected market conditions. This behavior shows that this formulation simply magnifies the direction of the market, performing incredibly well in a bull market and generating high expected returns, but conversely does poorly in a bear mark. 

This behavior is a direct contrast to the behavior of the pure directional formulation (figure \ref{fig:DirVNaive}) Where the trend line has a more gradual change with a smaller slope than the baseline. This shows that while the directional trading policy does not capitalize on opportunities to generate cash in a bull market and may perform worse than simply holding the portfolio, it is extremely good at minimizing losses in bad market conditions, a trait that is extremely helpful in recessionary conditions. 
\subsection{Trading Cost Analysis}
\begin{figure}[H]
    \centering
\begin{tabular}{lrrrrrrrrrr}
\toprule
{} & \multicolumn{5}{l}{\textbf{Number of Trades Executed}} & \multicolumn{5}{l}{\textbf{Trading Cost Incurred}} \\
{} &          Mean & Std. Dev. & Median &    Max &   Min &                  Mean & Std. Dev. & Median &     Max &   Min \\
policy      &               &           &        &        &       &                       &           &        &         &       \\
\midrule
DirP\_0      &         54.76 &     22.26 &   52.0 &  113.0 &  19.0 &                306.85 &    180.54 &  277.5 &  1017.0 &  42.0 \\
DirP\_25.0   &         46.83 &     18.16 &   44.5 &  104.0 &  12.0 &                263.79 &    151.25 &  232.0 &   832.0 &  48.0 \\
DirP\_50.0   &         39.10 &     13.17 &   36.0 &  106.0 &  12.0 &                219.65 &    116.61 &  204.0 &   630.0 &  42.0 \\
DirP\_75.0   &         38.81 &     11.79 &   36.5 &   78.0 &  12.0 &                217.62 &    110.39 &  207.0 &   621.0 &  44.0 \\
DirP\_500    &         38.74 &     11.63 &   36.5 &   69.0 &  12.0 &                217.76 &    111.59 &  208.5 &   621.0 &  46.0 \\
Directional &         20.22 &      7.06 &   20.0 &   42.0 &   6.0 &                112.76 &     60.10 &  102.0 &   288.0 &  20.0 \\
Naive       &         27.16 &      8.40 &   26.0 &   45.0 &   0.0 &                153.76 &     81.13 &  144.0 &   387.0 &   0.0 \\
\bottomrule
\end{tabular}

    \caption{Summary Statistics on Trading Behavior}
    \label{fig:Trade Behavior}
\end{figure}

When examining the number of executed trades and the corresponding incurred trading cost, we can see that the directional formulation significantly reduced the number of trades needed to achieve our target portfolio compared to our naive baseline and as a result the average amount of money wasted on trading fees is also significantly lower. In contrast, the directional penalty formulations as well as the unpenalized day trading variant all had much higher trading volumes with the day trading and $\lambda = 25\%$ variants trading more than twice the amount of the directional formulation. 

One can also see that as the penalty parameter is increased, the number of trades reduces as well, reinforcing $\lambda$ as a value to control accepted risk and aggressiveness. Additionally, the significant reduction in trading costs once again highlights the directional formulation's benefits of being a method to easily navigate a portfolio to a target while consistently minimizing lost value.

\subsection{Runtime Analysis}

As this is an approach designed for retail traders, in practice this would probably be delivered as a web-based application and as such, algorithmic runtime is important for both a seamless user experience and minimizing the cost of mounting the application on cloud platform that often charge on a usage basis. 

\begin{figure}[H]
    \centering
\begin{tabular}{lrrrrr}
\toprule
     policy &  Mean &  Std. Dev. &  Median &    Max &   Min \\
\midrule
      Naive & 10.43 &       4.37 &   10.31 &  24.11 &  3.22 \\
Directional & 22.26 &      13.05 &   20.17 & 166.39 &  7.77 \\
DirP\_500 & 33.87 &      16.17 &   30.70 & 104.45 & 11.67 \\
  DirP\_75.0 & 40.75 &      27.14 &   33.91 & 293.39 & 11.63 \\
  DirP\_50.0 & 42.28 &      29.70 &   35.23 & 295.14 & 11.51 \\
  DirP\_25.0 & 38.44 &      24.47 &   32.93 & 222.33 & 11.48 \\
     DirP\_0 & 33.80 &      14.56 &   31.63 &  96.73 & 12.67 \\
\bottomrule
\end{tabular}

    \caption{Summary Statistics on Total Algorithm Runtime (s)}
    \label{fig:RuntimeTable}
\end{figure}

Figure \ref{fig:RuntimeTable} shows the total runtime taken by the optimization policies over the simulation. While in practice this optimization would be done over 30 days, these results are helpful for showcasing the performance of our policies. While the total runtime across the simulation is significantly higher than the naive formulation, in practice these differences are negligible and on average the optimization would take 1-2 seconds per day with the directional policy taking even less time.   

\begin{figure}[H]
    \centering
    \includegraphics[width=0.5\textwidth]{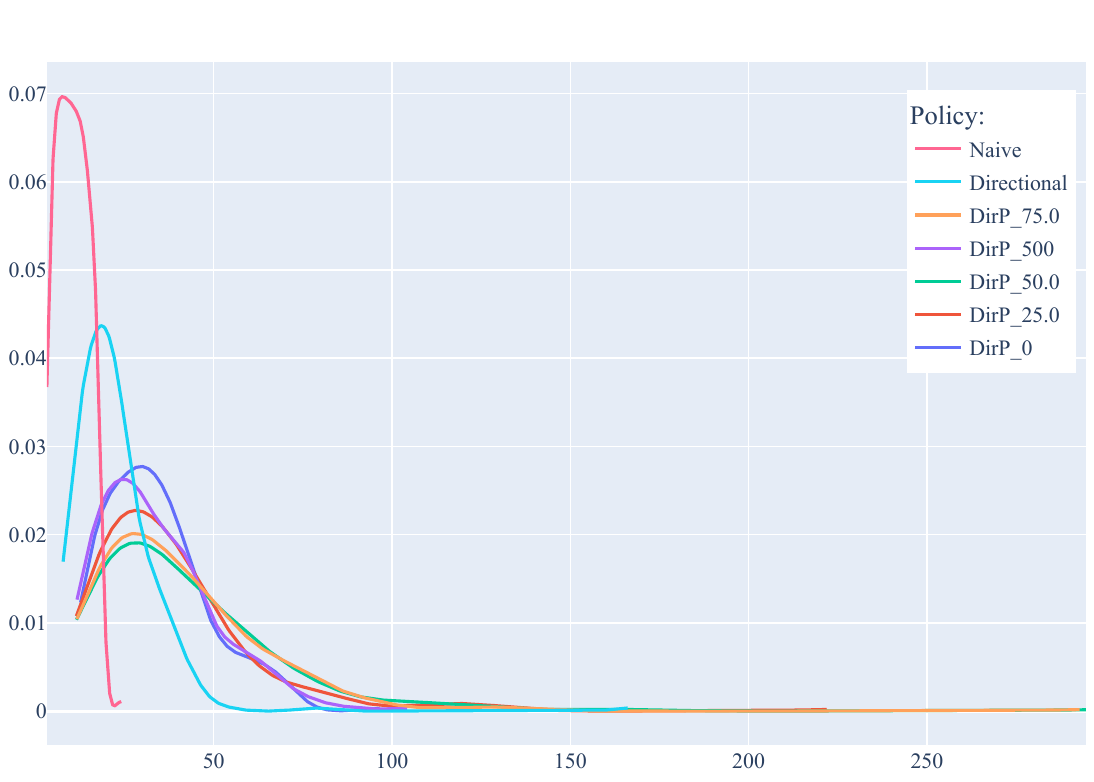}
    \caption{Probability Distribution of the runtimes}
    \label{fig:runtimehist}
\end{figure}

As seen in figure \ref{fig:runtimehist}, the distribution of the runtimes has an extremely long tail and the difference between the median and the max runtimes for all our policies is extremely high. Additionally, while the median and mean runtimes for the penalty policies are similar, the length of the tails are drastically different. 

An interesting observation is that the constrained directional policy ran in almost half the time as the penalized policies, despite the approach of the penalized policy being similar to taking a Lagrangian relaxation of the constraints \ref{buy_restrict} and \ref{sell_restrict}. One potential reason for this runtime difference is that these two constraints trim a large portion of the feasible space in turn reducing the maximum possible size of the branch-and-bound tree. Additionally, the average runtime does not seem to correlate with penalty values as having a $\lambda = 0$ has an extremely similar runtime distribution to an extremely high penalty.

\section{Column Generation}
\subsection{Introduction \& Methodology}
Column Generation is a method for solving complex IPs through feasible solution enumeration. It leverages the structure of a Dantzig reformulation to enumerate feasible solutions based on patterns in the problem constraints. In our problem, such patterns are present in the sell and buy behavior. We hope to lean on the enumeration idea of column generation to leverage the structure embedded in the trading pattern.

As a reminder, the problem we are solving involves a portfolio turnover from a current portfolio to a target portfolio in a defined number of days. The algorithm must determine the timing of the trades that would make a liquidation or purchase of positions most optimal, namely, by preserving portfolio value against transaction costs and market fluctuations. Since we have the target portfolio's construction, the quantity bought or sold is assumed to be the quantity that satisfies the difference between the former and the initial portfolio - no more, no less. In other words, knowing how much is sold is partly irrelevant to the trader since it's assumed that it's the necessary quantity. In addition, the trader would not be asked to buy or sell the same asset more than once as that would defeat the goal of finding the ``best'' time to sell/buy. As such, we can apply a Dantzig reformulation, if we consider the enumerated feasible solutions to be the time period at which an action is taken for each asset in the portfolio (Formulation \ref{eq:cg-form}). We expose the Dantzig reformulation below.

First, we have:
\begin{align}
\underset{z,w}{\text{maximize}}&\ C_\mathbf{T} +  Y_{\mathbf{T}}^TP_\mathbf{T} - \sum_{\tau = t}^\mathbf{T}\sum_{a \in A} F(w^+_{\tau, a} + w^-_{\tau,a})&\ \\
\text{subject to}&\ P_{\tau} = P_{\tau-1} + z^+_\tau - z^-_\tau &\ \forall \tau = t+1, \ldots, \mathbf{T}\\ 
&\ C_\tau = C_{\tau-1} - Y_\tau^T(z_\tau^+ - z_\tau^-) - \sum_{a\in A}F(w^+_{\tau,a} + w^-_{\tau,a}) &\ \forall \tau =  t+1, \ldots, \mathbf{T}\\ 
&\ z^+_\tau \leq M_\tau w^+_{l,\tau} &\ \forall \tau = t, t+1, \ldots, \mathbf{T} \\
&\ z^-_\tau \leq M_\tau w^-_{k,\tau} &\ \forall \tau = t, t+1, \ldots, \mathbf{T}  \\
&\ P_\tau \geq \Vec{0} &\ \forall \tau = t, t+1, \ldots, \mathbf{T} \\ 
&\ C_\tau \geq 0 &\ \forall \tau = t, t+1, \ldots, \mathbf{T} \\ 
&\ P_\mathbf{T} \geq \mathcal{T} \\
&\ z_\tau^+, z_\tau^- \in \mathbb{Z}_+^{|A|} &\ \forall \tau = t, t+1, \ldots, \mathbf{T} \\
&\ w^+_{l,\tau} \in W^+_{\tau}\ &\ \forall \tau = t, t+1, \ldots, \mathbf{T} \\
&\ w^-_{k,\tau} \in W^-_{\tau}\ &\ \forall \tau = t, t+1, \ldots, \mathbf{T} \\
\end{align}
With,

\begin{align}
    W^+ = \{w^+_l \in \{0,1\}^{[T\text{x}N]}: w^+_l + w^-_l \leq 1, \sum^T_{\tau}w^+_{l,\tau} \leq 1\} \\
    W^- = \{w^-_k \in \{0,1\}^{[T\text{x}N]}: w^+_k + w^-_k \leq 1, \sum^T_{\tau}w^-_{k,\tau} \leq 1\}
\end{align}
Which gives us the following Dantzig formulation:
\begin{align}
\underset{z,\lambda, \gamma}{\text{maximize}}&\ C_\mathbf{T} +  Y_{\mathbf{T}}^TP_\mathbf{T} - \sum_{\tau = t}^\mathbf{T}\sum_{a \in A} F(w^+_{\tau, a} + w^-_{\tau,a})&\ \label{eq:cg-form}\\
\text{subject to}&\ P_{\tau} = P_{\tau-1} + z^+_\tau - z^-_\tau &\ \forall \tau = t+1, \ldots, \mathbf{T} \\ 
&\ C_\tau = C_{\tau-1} - Y_\tau^T(z_\tau^+ - z_\tau^-) - \sum_{a\in A}F(w^+_{\tau,a} + w^-_{\tau,a}) &\ \forall \tau =  t+1, \ldots, \mathbf{T}\\ 
&\ z^+_\tau \leq M_\tau w^+_{l,\tau} &\ \forall \tau = t, t+1, \ldots, \mathbf{T} \\
&\ z^-_\tau \leq M_\tau w^-_{k,\tau} &\ \forall \tau = t, t+1, \ldots, \mathbf{T}  \\
&\ P_\tau \geq \Vec{0} &\ \forall \tau = t, t+1, \ldots, \mathbf{T} \\ 
&\ C_\tau \geq 0 &\ \forall \tau = t, t+1, \ldots, \mathbf{T} \\ 
&\ P_\mathbf{T} \geq \mathcal{T} \\
&\ w^+_{\tau} = \lambda_l \cdot w^+_{l, \tau} &\ \forall \tau = t, t+1, \ldots, \mathbf{T}\\
&\ w^-_{\tau} = \gamma_k \cdot w^-_{k, \tau} &\ \forall \tau = t, t+1, \ldots, \mathbf{T}\\
&\ \sum^L_{L=1} \lambda_{l} = 1 \\
&\ \sum^K_{k=1} \gamma_{k} = 1 \\
&\ z_\tau^+, z_\tau^- \in \mathbb{Z}_+^{|A|} &\ \forall \tau = t, t+1, \ldots, \mathbf{T} \\
&\ \lambda_l \in \{0,1\}\ & \forall l=1, \ldots, L \\
&\ \gamma_k \in \{0,1\}\ & \forall k=1, \ldots, K
\end{align}
In plain terms, the sets \(W^+,\ W^-\), are both respective enumerations of all the buy actions and sell actions (when to buy and when to sell). For an optimization with 3 assets, 30 lookahead days and 2 assets needing to be liquidated, the \(W^-\) set will have (including: ``no sell action as a possible action'') \(30*30+2*30+1 = 961\) different possible combinations of sell actions. This value grows exponentially with the number of assets and lookahead periods. The key insight is that the number of possible options dramatically decreases once an asset reaches its target state. In fact, if this occurs before the trading horizon is complete, no more feasible solution will be generated for this asset, i.e, \(\gamma = \Vec{0}\). 
\subsection{Results}
We implemented two versions of our column generation (CG) algorithm. The first one (ColGen\_False) only enumerated the possible buy strategies and maintained the sell actions as defined in formulation \ref{eqn:gen_obj}. The second version (ColGen\_True) includes the enumeration of both buy and sell strategies. Both were compared for the sake of strategic and numerical performance.
\begin{table}[h]
    \centering
    \begin{tabular}{lrrrrr}
        \toprule
              policy &  Mean &  Std. Dev. &  Median &   Max &    Min \\
        \midrule
               Naive &  3.16 &       6.37 &    5.46 &  8.98 &  -7.19 \\
         Directional &  3.08 &       6.31 &    5.46 &  8.99 &  -7.19 \\
           DirP\_25.0 &  3.08 &       6.31 &    5.46 &  8.99 &  -7.19 \\
              DirP\_0 &  8.43 &      12.56 &    8.73 & 19.27 & -12.11 \\
         ColGen\_True &  3.11 &       6.34 &    5.46 &  8.99 &  -7.20 \\
        ColGen\_False &  3.11 &       6.33 &    5.46 &  8.98 &  -7.19 \\
        \bottomrule
    \end{tabular}
    \caption{Summary Statistics on the Percent Change (\%) in Portfolio Value}
    \label{tab:GainStatsCG}
\end{table}
\begin{table}[h]
    \centering
    \begin{tabular}{lrrrrrrrrrr}
        \toprule
        {} & \multicolumn{5}{l}{Amount Traded} & \multicolumn{5}{l}{Trading Cost Incurred} \\
        {} &          Mean & Std. Dev. & Median &   Max &  Min &                  Mean & Std. Dev. & Median &    Max &   Min \\
        policy       &               &           &        &       &      &                       &           &        &        &       \\
        \midrule
        ColGen\_False &           2.0 &      1.41 &    3.0 &   3.0 &  0.0 &                   7.0 &      4.64 &    9.0 &   12.0 &   0.0 \\
        ColGen\_True  &           2.2 &      1.30 &    3.0 &   3.0 &  0.0 &                   8.0 &      4.64 &    9.0 &   12.0 &   0.0 \\
        DirP\_0       &          14.0 &      7.68 &   18.0 &  21.0 &  3.0 &                  55.8 &     34.77 &   57.0 &  108.0 &  15.0 \\
        DirP\_25.0    &           2.6 &      1.14 &    3.0 &   4.0 &  1.0 &                  10.2 &      4.09 &    9.0 &   16.0 &   5.0 \\
        Directional  &           1.8 &      1.30 &    2.0 &   3.0 &  0.0 &                   6.4 &      4.51 &    6.0 &   12.0 &   0.0 \\
        Naive        &           2.4 &      0.89 &    3.0 &   3.0 &  1.0 &                   9.4 &      2.88 &    9.0 &   12.0 &   5.0 \\
        \bottomrule
    \end{tabular}
    \caption{Summary Statistics on Trading Behavior}
    \label{tab:TradeBehaviorCG}
\end{table}
\begin{table}[h]
    \centering
    \begin{tabular}{lrrrrr}
        \toprule
              policy &  Mean &  Std. Dev. &  Median &   Max &   Min \\
        \midrule
               Naive &  3.39 &       0.12 &    3.38 &  3.58 &  3.27 \\
         Directional &  9.55 &       0.17 &    9.60 &  9.71 &  9.32 \\
           DirP\_25.0 & 10.00 &       0.17 &    9.96 & 10.28 &  9.85 \\
              DirP\_0 & 11.26 &       0.63 &   11.24 & 12.02 & 10.64 \\
         ColGen\_True &  9.32 &       2.79 &    9.35 & 13.79 &  6.57 \\
        ColGen\_False &  9.55 &       0.17 &    9.53 &  9.75 &  9.31 \\
        \bottomrule
    \end{tabular}
    \caption{Summary Statistics on Total Algorithm Runtime (s)}
    \label{tab:RuntimeTableCG}
\end{table}

The CG strategies performed almost identically on all metrics (Table \ref{tab:GainStatsCG}). The algorithm achieves a better mean return than the strict directional algorithm with minimal additional variance. Further, similar performance is observed when we consider the number of trades and the amount of trading costs incurred (Table \ref{tab:TradeBehaviorCG}). On all accounts, the CG methods, place themselves between the performance of the Naive method and the strict directional method.

We note that the solve times of the CG enumeration methods are on average less than their directional counterpart, but, they suffer from large variances. This is the direct result of large number of enumerations at the beginning of the trading horizon  (Table \ref{tab:RuntimeTableCG}).

\subsection{Discussion}
In general, the performance of the enumeration method is not particularly advantageous though it does beat the directional approach. Where this method becomes beneficial is when trades are completed before the end of the trading horizon. Then, the optimization is systematically constrained to stop considering possible trades for a given asset.

We realize that our algorithm could benefit from a complete implementation of the CG algorithm. Instead of generating and optimizing over all possible trade patterns we could generate subsets of feasible solutions to create interesting bounds to be utilized in a branch-and-bound algorithm. Unfortunately, with limited time, we will have to leave this out of the scope of this project. In the same vein, we believe there is space to explore intelligent generation of subsets that would allow the BnB algorithm to quickly identify a tight bound to the problem which could then serve to easily solve the MIP problem. For example, one could divide subsets based on chunks of days. More work needs to be undertaken to develop such an approach. 
\section{Conclusion}
In this work, we present a multi-period receding horizon framework along with two different integer programming based optimization policies that that allow retail traders to efficiently transition the holdings in their current portfolio to one that satisfies their investment targets. The first policy, also known as the directional trading policy, limits trades to only ones that make progress in the transition and leverages price volatility to identify trading moments that minimize the trades needed. The other policy, also known as the directional penalty policy, allows the user to specify their acceptable risk and suggests trades that maximize the end portfolio value. Experimental results show that the directional trading policy significantly reduces the amount of money wasted on trading fees and reduces losses to portfolio value in bear markets when compared to a single-period transition. The directional penalty policy also shows significant potential and allows the user to achieve significant returns in bull markets when users are willing to accept higher risk and moderate gains in portfolio value when traders want lower risks. Additionally the computational load of all our policies is relatively low, allowing our methodologies to be easily implemented in practice, potentially on low-cost cloud computing platforms. These promising results show the potential of integer programming as a tool to help retail investors better manage their portfolios. One shortcoming of this work is the assumptions that the forecast will have reasonable error. 

Black swan events can occur and can result in extremely inaccurate forecasts, a case we do not explicitly handle. While a receding horizon framework slightly mitigates these risks, including some explicit risk-handling measures could greatly improve our framework. Additionally, our framework would obviously benefit from high quality forecasts therefore anyone aiming to implement the methodologies described in this paper for their own use should develop a strong forecasting algorithm they can trust. On a similar note, future directions of work for this problem include modifying the prediction models used to generate price estimates to one that generates probabilistic forecasts and leverage multi-stage stochastic programs as the optimization policy. While this would increase the computational load, approaching the portfolio transition problem through this lens can allow for safer, more profitable decision making. 

Another potential avenue for exploration is splitting up the problem into two-stages, one to decide when to trade and another to decide how much to trade on the days decided. This reformulation combined with the multi-stage receding horizon framework could allow for significant runtime improvements by not solving the second-stage problem if the number of trades decided for a given day is zero, as only the trade recommendations for the current day is accepted in our framework.  

The nature of the portfolio transition problem also lends itself nicely to a Smart "Predict, then Optimize" approach such as the one proposed by \cite{elmachtoub2022smart}. Instead of having our forecasting and optimization as two separate components in our work flow, recent work has made it possible to integrate optimization models directly into the prediction workflow and have a neural network directly predict the optimization output \citep{tang2022pyepo}. 

Along with new techniques, further improvements could be made to our work by including other components of trading such as dynamic trade pricing, dividends, bid-ask spreads, and trade latency.

\bibliographystyle{unsrtnat}
\bibliography{references}  
\end{document}